\documentclass[12pt,osajnl2,preprint,showpacs]{revtex4}  %% REVTeX 4.0

\usepackage[draft]{hyperref} %% optional
\usepackage{graphicx}% Include figure files
\usepackage{subfigure}
\usepackage{subeqn}

\begin{document}

\title{(2+1)D surface solitons at the interface between a linear medium and a nonlocal nonlinear medium}

%% For REVTeX it is possible to automate superscript and e-mail callouts with the superscriptaddress option; see REVTeX4 documentation.

\author{Zhiwei Shi$^{1,2}$, Huagang Li$^3$ and Qi Guo$^{1,*}$}
\address{$^1$Key Laboratory of Photonic Information
Technology of Guangdong Higher Educaiton Institutes, South
China Normal University, \\ Guangzhou 510631, P.R.China}
\address{$^2$Faculty of Information Engineering, Guangdong
University of Technology, \\ Guangzhou 510006,
P.R.China}
\address{$^3$Department of Physics, Guangdong Institute of Education, \\Guangzhou 510303, P.R.China}
\address{$^*$Corresponding author: guoq@scnu.eud.cn}

\begin{abstract}We address (2+1)D surface solitons occurring at the interface between a linear medium and a nonlocal nonlinear medium whose nonlinear contribution to the refractive index has a initial value at the interface.
We find that there exist stable single and dipole surface solitons which do not exhibit a power threshold. The properties of the surface solitons can be affected by the initial value and the degree of nonlocality.
When a laser beam is launched away from the interface, the beam will be periodic oscillations.
\end{abstract}

%\ocis{190.4350, 190.6135, 240.6990.}% REPLACE WITH CORRECT OCIS CODES FOR YOUR ARTICLE
                          % NOTE: \ocis{} IS ALIASED TO \pacs{} BUT MUST
                          % FORMAT THE TERMS CORRECTLY FOR EACH JOURNAL

\maketitle %% null function with osajnl.sty

\section{Introduction}
Nonlocal spatial solitons have been
investigated for decades both theoretically and experimentally. Spatially nonlocality is a generic property in different materials including the photorefractive media~\cite{ref1,ref2}, thermal nonlinear media~\cite{ref3,ref4,ref5,ref6,ref7}, liquid crystals~\cite{ref8,ref9,ref10,ref11,ref12,ref12-1}, and so on.
Nonlocality can lead to new kinds of waves that would have been otherwise impossible in local nonlinear media. Especially, for two-dimensional media with different types of nonlocal
response, they can support stationary multipoles~\cite{ref13,ref14,ref15,ref16,ref17,ref18}, stable
vortices~\cite{ref19,ref20,ref21,ref22,ref23}, and rotating~\cite{ref24,ref25} and spiraling~\cite{ref5,ref26} soliton states.

Surface waves localized at the interface of two different optical materials have many novel properties, which have been studied in nonlocal media recently.
The light beam trajectory can be strongly affected by the presence
of interfaces, because beams propagating in nonlocal media cause refractive index changes in regions far
exceeding the beam width. Under proper conditions, stationary surface waves can propagate along the interface in both local nonlinear media~\cite{ref27,ref28,ref29,ref30} and nonlocal nonlinear media~\cite{ref31,ref32,ref33,ref34,ref35,ref36,ref37,ref38}.

Thermal media~\cite{ref33,ref34,ref37} and photorefractive crystals~\cite{ref38} have been also utilized to demonstrate (2+1)D surface
solitons. In this paper, we will study (2+1)D surface solitons occurring at the interface between a linear medium and a nonlocal nonlinear medium whose nonlinear contribution to the refractive index has a initial value at the interface.
We find that there exist single surface solitons at the edge and the corner of the two 2D media and dipole surface solitons at the edge of the two 2D media. These stable solitons do not exhibit a power threshold. The positions of the peak values and full
width at half maximum(FWHM) of the surface solitons can be affected by the degree of nonlocality. However, the initial value can only influence the positions of the peak values of the surface solitons. In addition, when a laser beam is launched away from the interface, the beam will be periodic oscillations, even if the launch position is far away from
the interface.

\section{Theoretical Model}
Considering a laser beam propagating
along the interface between a nonlocal nonlinear medium and a linear medium, the complex
amplitude $E(X,Y,Z)$ of the light field satisfies the scalar
wave equation~\cite{ref31,ref33,ref38}
\begin{eqnarray}
\nabla_{XY}^{2} E+\frac{\partial^{2}E}{\partial Z^{2}}+k_0^2
n^{2}E=0, \label{eq:one}
\end{eqnarray}
where $k_{0}=2\pi/\lambda$ is the wave number in vacuum, and
$n=n_L+\Delta n$ for the nonlinear medium($X\leq 0$) and $n=n_0$ for
the linear medium($X\geq 0$). $\Delta n$ represents the
nonlinear contribution to the refractive index and may
originate from any diffusive nonlinear effect, which can be written by~\cite{ref31}
\begin{eqnarray}
w_{m}^{2}\nabla_{XY}^{2}\Delta n-\Delta
n+n_2|E|^{2}=0, \label{eq:two}
\end{eqnarray}
where $w_{m}$ is the characteristic length of the nonlinear
response and $n_2$ is the nonlinear index coefficient. For the local
case, $w_{m}\rightarrow0$, we have $\Delta n=n_2|E|^{2}$.

Let us put $E(X,Y,Z)=E_{0}(X,Y,Z)\exp(i\beta Z)$ and submit into the
equation (1) and (2). Then, using the slowly varying envelope
approximation and introducing the normalized variables $x=X/w_{0}, y=Y/w_{0}, z=Z/(\beta
w_{0}^{2}), a=k_{0}w_{0}\sqrt{n_{2}n_{L}}E_{0}$ and
$\phi=k_{0}^{2}w_{0}^{2}n_{L}\Delta n$, we get

\begin{subequations}
\label{eq:three}
\begin{equation}
\frac{1}{2}\nabla_{\bot}^{2}a+i\partial_{z}a+\beta_{1}a+\phi
a=0 \quad
\textrm{for}\quad x\leq 0,\label{subeq:3a}
\end{equation}
\begin{equation}
\frac{1}{2}\nabla_{\bot}^{2}a+i\partial_{z}a+\beta_{2}a=0 \quad
\textrm{for}\quad x\geq 0,\label{subeq:3b}
\end{equation}
\end{subequations}
and
\begin{eqnarray}
d^{2}\nabla_{\bot}^{2}\phi-\phi+|a|^{2}=0 \quad
\textrm{for}\quad x\leq 0,\label{eq:four}
\end{eqnarray}
where $\nabla_{\bot}^{2}=\partial_{x}^{2}+\partial_{y}^{2}$, $\beta_{1}=w_{0}^{2}(k_{0}^{2}n_{L}^{2}-\beta^{2})/2$,
$\beta_{2}=w_{0}^{2}(k_{0}^{2}n_{0}^{2}-\beta^{2})/2$, $\beta$ is the wave number in
the media, $w_{0}$
is the beam width, and $d=w_{m}/w_{0}$ stands for the degree of nonlocality
of the nonlinear response. For $x\leq 0$, the
equations describe a local nonlinear response as
$d\rightarrow0$ and a strongly nonlocal response as
$d\rightarrow\infty$.

We search for stationary soliton solutions of Eqs.~(\ref{eq:three})
and (\ref{eq:four}) numerically in the form $a(x,y,z)=u(x,y)\exp(i b
z)$, where $u$ is the real function and $b$ is a real propagation
constant of spatial solitons in the normalized system.

\begin{subequations}
\label{five}
\begin{equation}
\frac{1}{2}\nabla_{\bot}^{2}u-bu+\beta_{1}u+\phi
u=0 \quad
\textrm{for}\quad x\leq 0,\label{subeq:5a}
\end{equation}
\begin{equation}
\frac{1}{2}\nabla_{\bot}^{2}u-bu+\beta_{2}u=0 \quad
\textrm{for}\quad x\geq 0,\label{subeq:5b}
\end{equation}
\end{subequations}
and
\begin{eqnarray}\label{eq:six}
d^{2}\nabla_{\bot}^{2}\phi-\phi+|u|^{2}=0 \quad
\textrm{for}\quad x\leq 0.
\end{eqnarray}

\section{Numerical results}

\subsection{Single surface solitons}
We firstly talk about new surface-wave soliton solutions at the edge of the two 2D media. Here, we assume that the normalized
nonlinear contribution to the refractive index at the
interface($y=0$) has a initial value $\phi_{d}(\phi_{d}>0)$. The boundary
conditions for the fields at the interface are the continuity of the
transverse field($a(y\rightarrow0_{-})=a(y\rightarrow0_{+})$) and
its derivative($da(y\rightarrow0_{-})/dy=da(y\rightarrow0_{+})/dy$). Because the width
of the surface solitons is much smaller than the sample width, $a$ and $\phi$
vanishes at the other boundary.

\begin{figure}[htbp]
\centering
{\includegraphics[width=4.5cm,height=4cm]{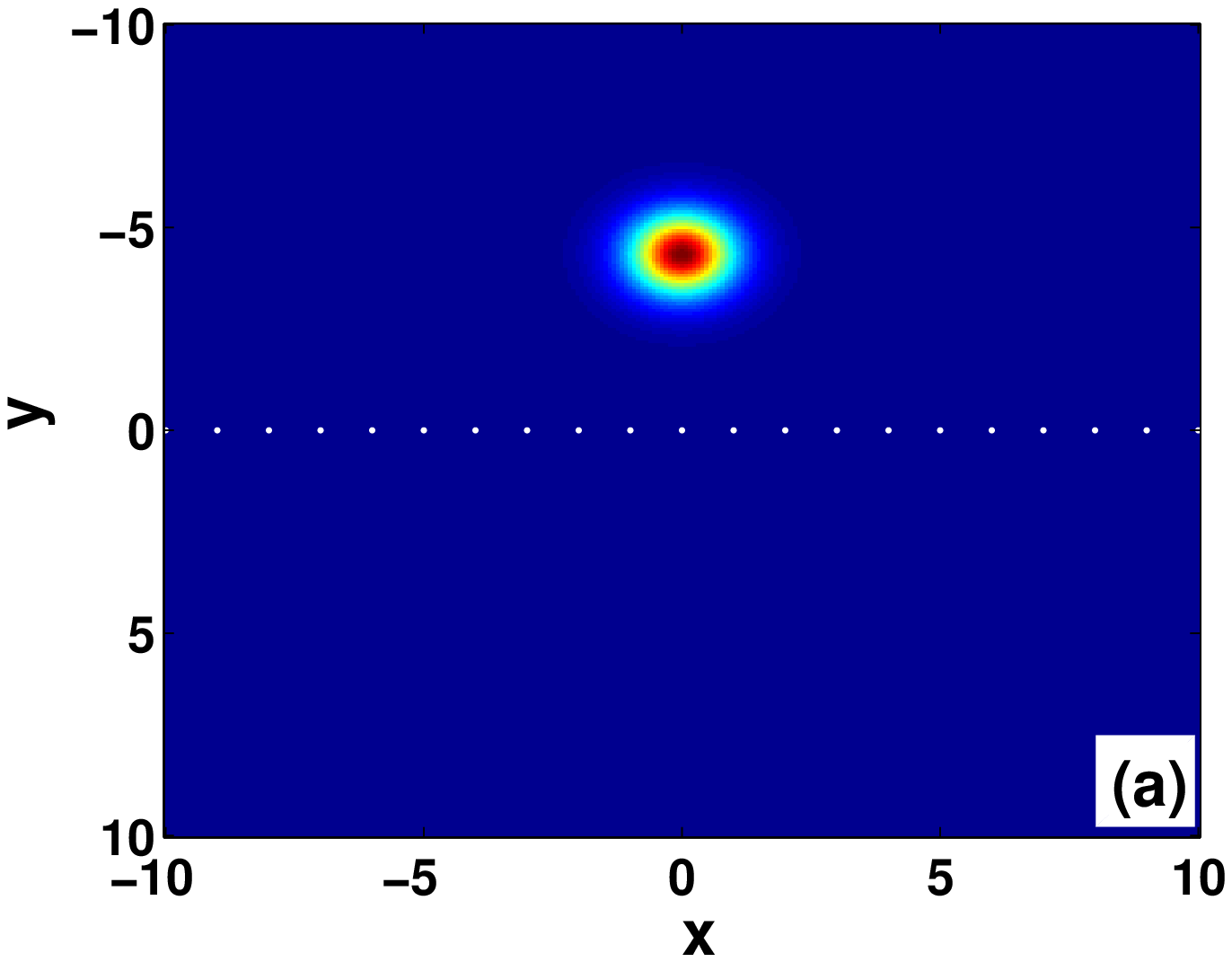}}
{\includegraphics[width=4.5cm,height=4cm]{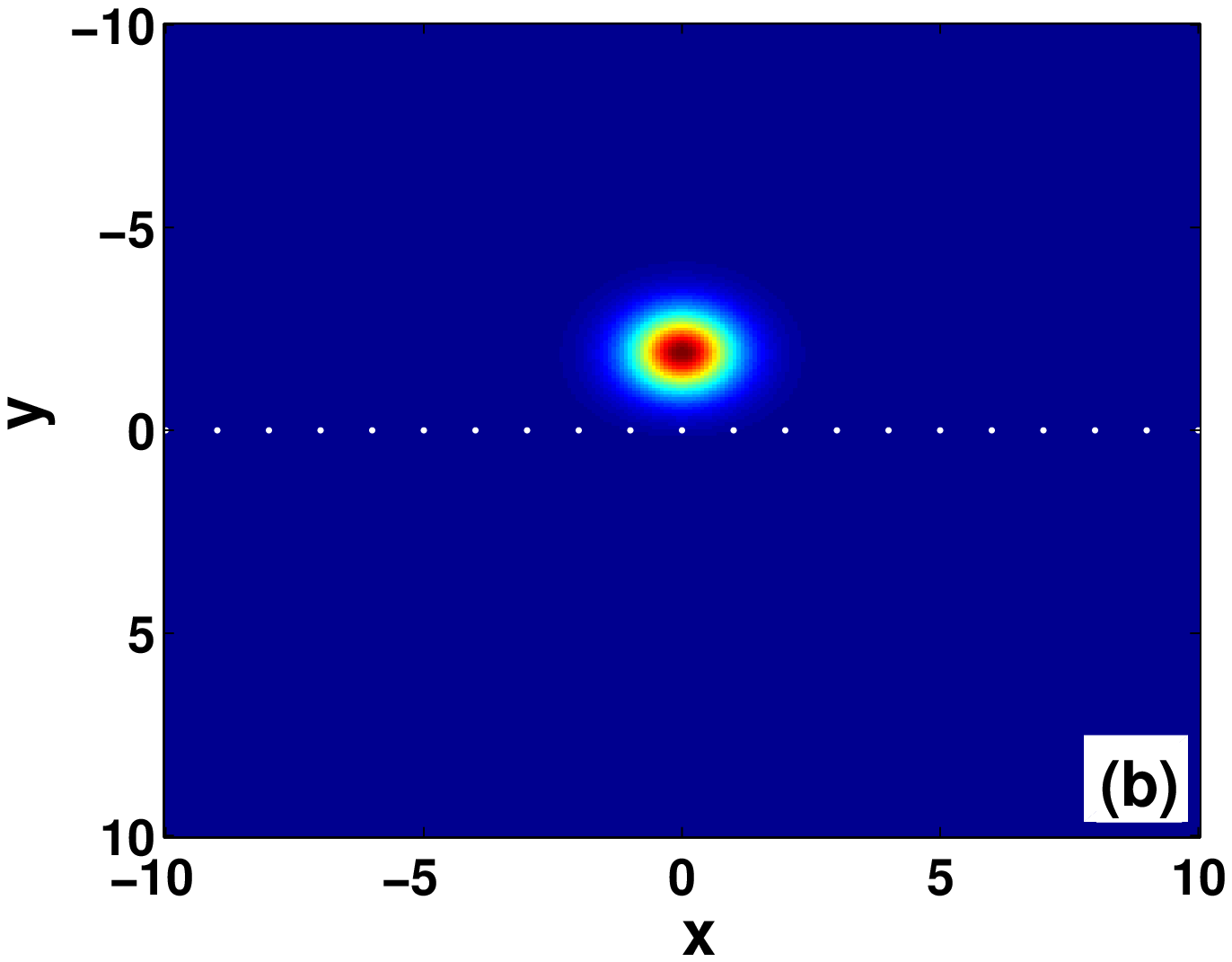}}
{\includegraphics[width=4.5cm,height=4cm]{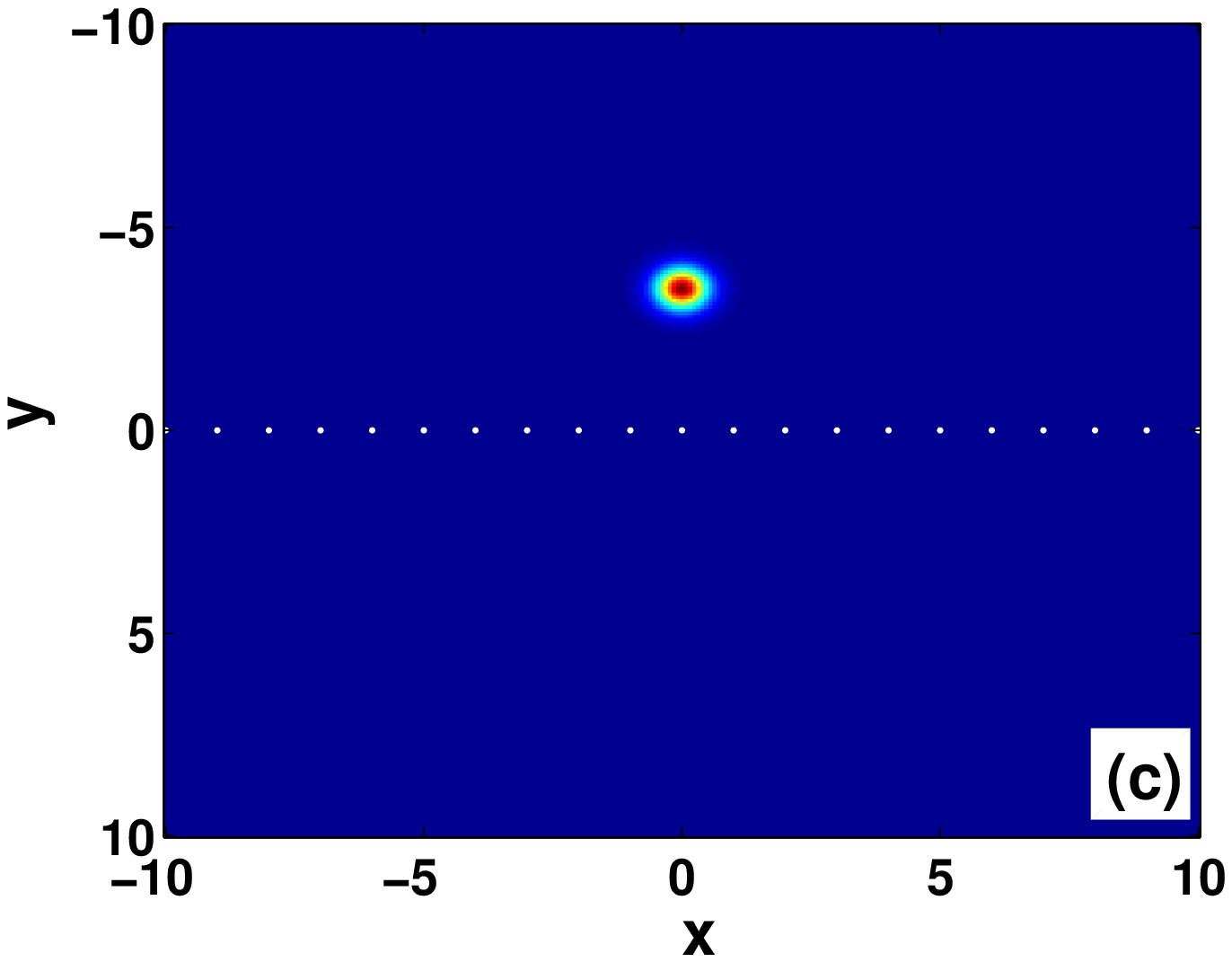}}
{\includegraphics[width=4.5cm,height=4cm]{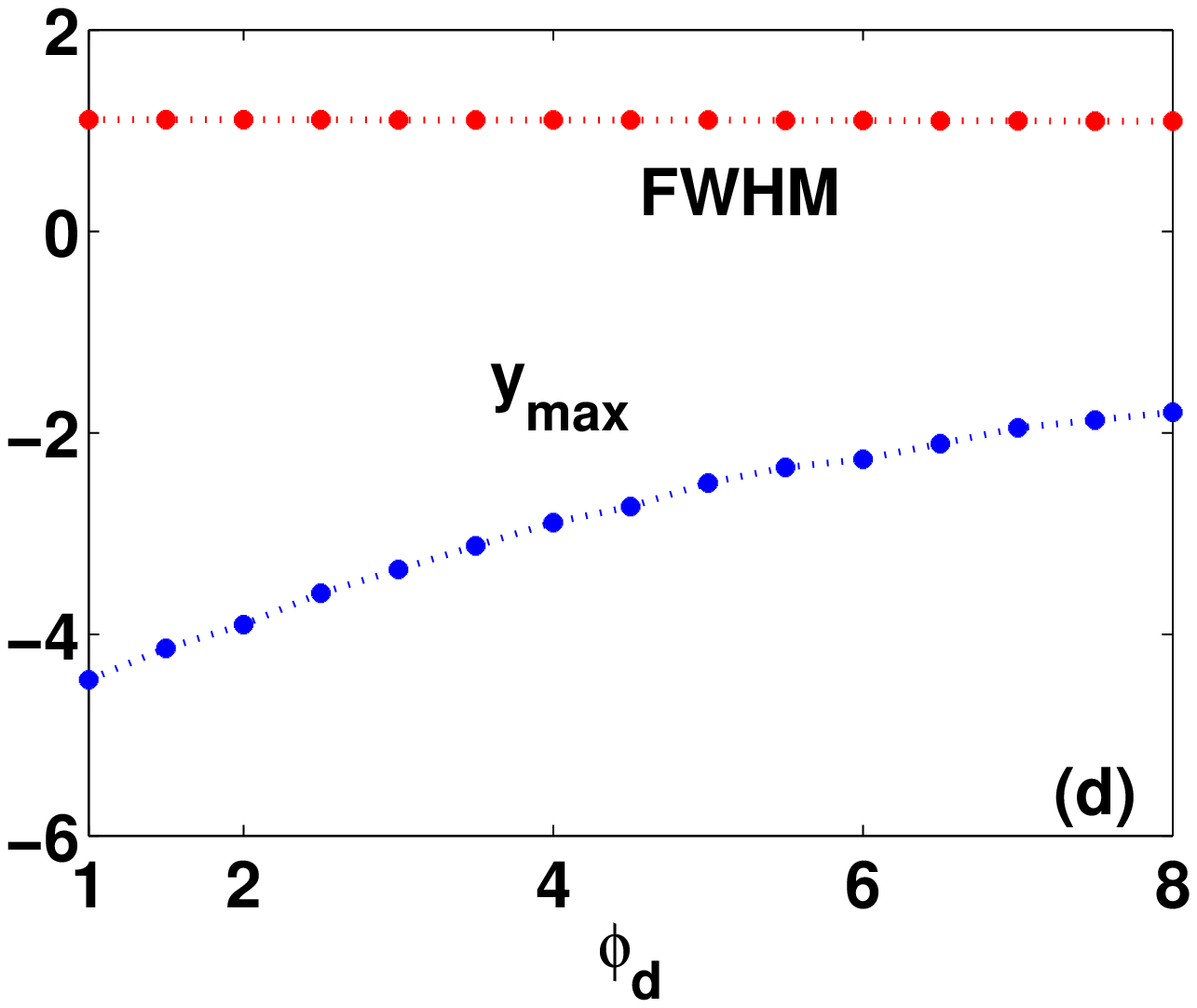}}
{\includegraphics[width=4.5cm,height=4cm]{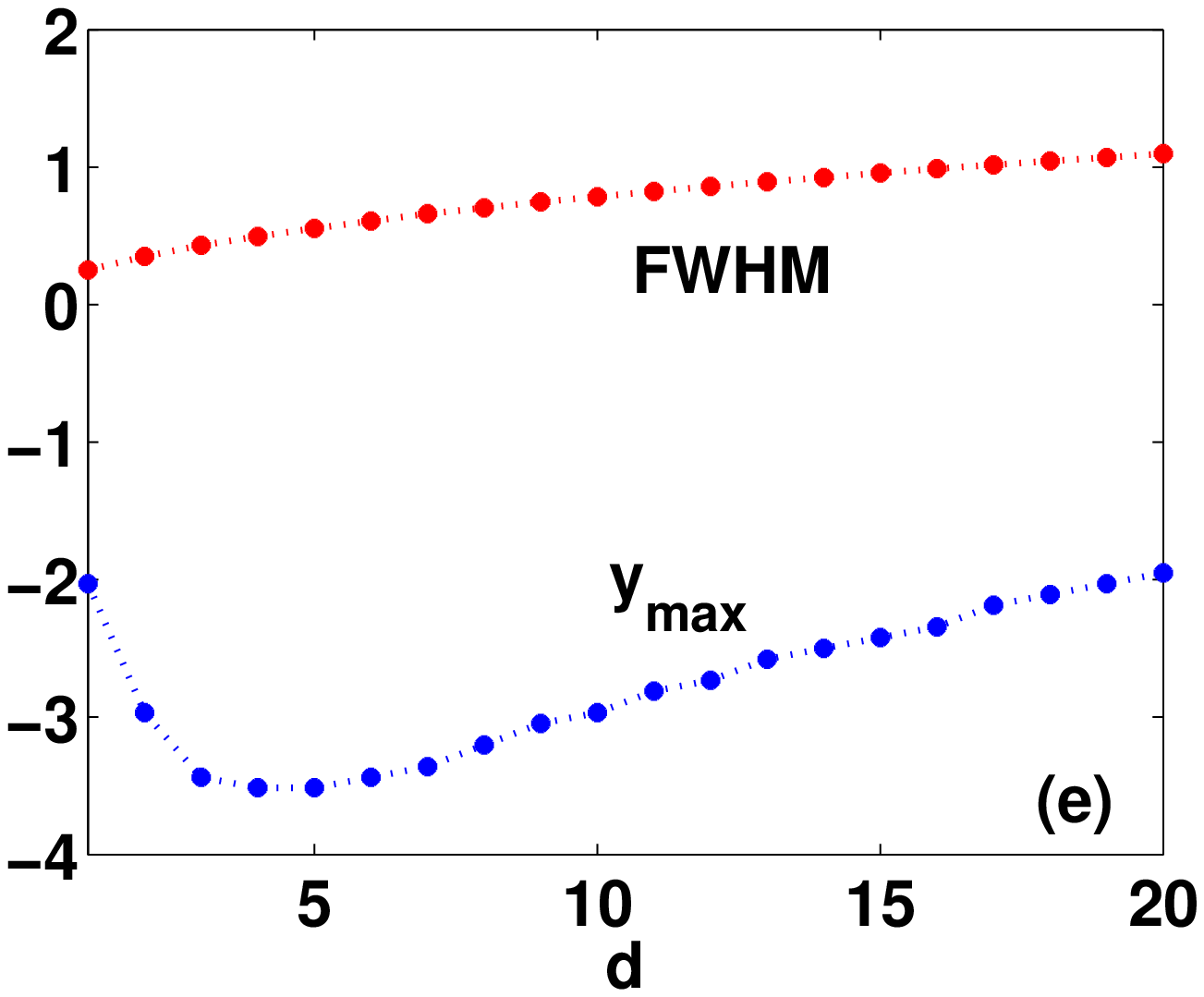}}
\caption{Sketch of 2D single surface solitons at the edge of the interface with (a) $\phi_{d}=1$, $d=20$, (b)
$\phi_{d}=7$, $d=20$ and (c) $\phi_{d}=7$, $d=5$. The positions of the peak values $y_{max}$ and FWHM versus
(d) the boundary value $\phi_{d}$ and (e) the
nonlocal degree $d$. White dashed line indicates interface position. All quantities are plotted in arbitrary dimensionless units.} \label{fig:one}
\end{figure}

Three different solitons are separately shown in
Fig.~\ref{fig:one}(a), (b) and (c). We can easily see that the
soliton is closely attached to the interface with $\phi_{d}=7$ and
$d=20$ in Fig.~\ref{fig:one}(b), but the solitons are farther
detached from the interface with $\phi_{d}=1$ and $d=20$ in
Fig.~\ref{fig:one}(a) or $\phi_{d}=7$ and $d=5$ in
Fig.~\ref{fig:one}(c). Obviously, $\phi_{d}$ and $d$ influence the position or shape of the solitons.
To further explain this point, we see that the positions of the peak values($y_{max}$) and FWHM of the surface wave solitons versus the boundary value $\phi_{d}$(Fig.~\ref{fig:one}(d)) at the interface
or the degree of nonlocality
$d$(Fig.~\ref{fig:one}(e)). From Fig.~\ref{fig:one}(d), we can see that, under the condition($d=20$), the soliton will
be attracted to the interface and more and more significant part of their optical
power residing in the linear meidium as $\phi_{d}$ increases. That is to say, $\phi_{d}$ is larger, the larger a ``surface force"
exerted on the beam by the interface. However, $\phi_{d}$ cannot influence the beam width of solitons. From Fig.~\ref{fig:one}(e), at $\phi_{d}=7$, the changing of the refractive index of the nonlocal nonlinear medium induced by $d$ results in the changing of FWHM and $y_{max}$ of the solitons. Because of the changing of FWHM, $y_{max}$ changes intricately, though a force exerted on the beam by the degree of nonlocality increases all the while.

\begin{figure}[htbp]
\centering
{\includegraphics[width=4.5cm,height=4cm]{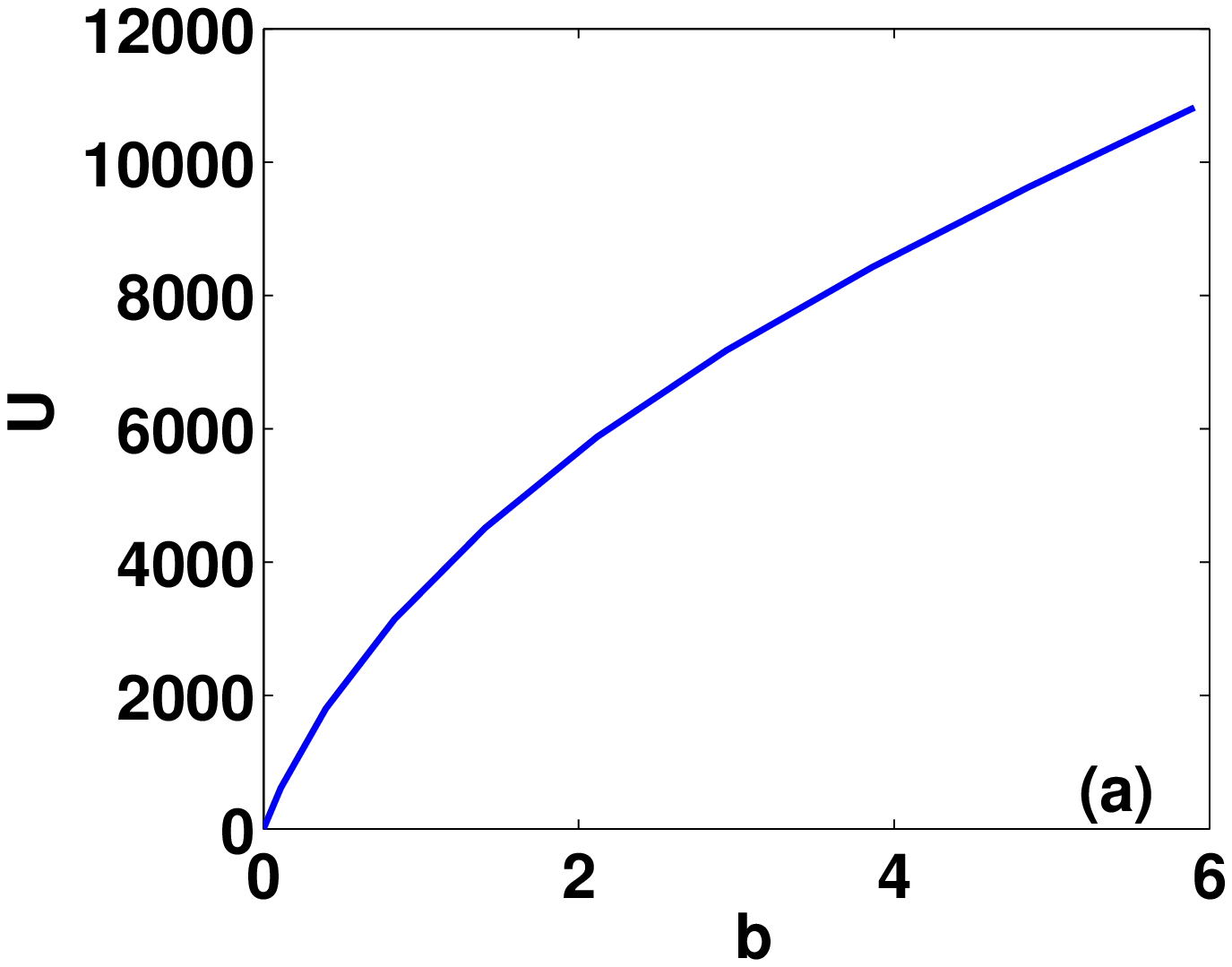}}
{\includegraphics[width=6cm,height=5cm]{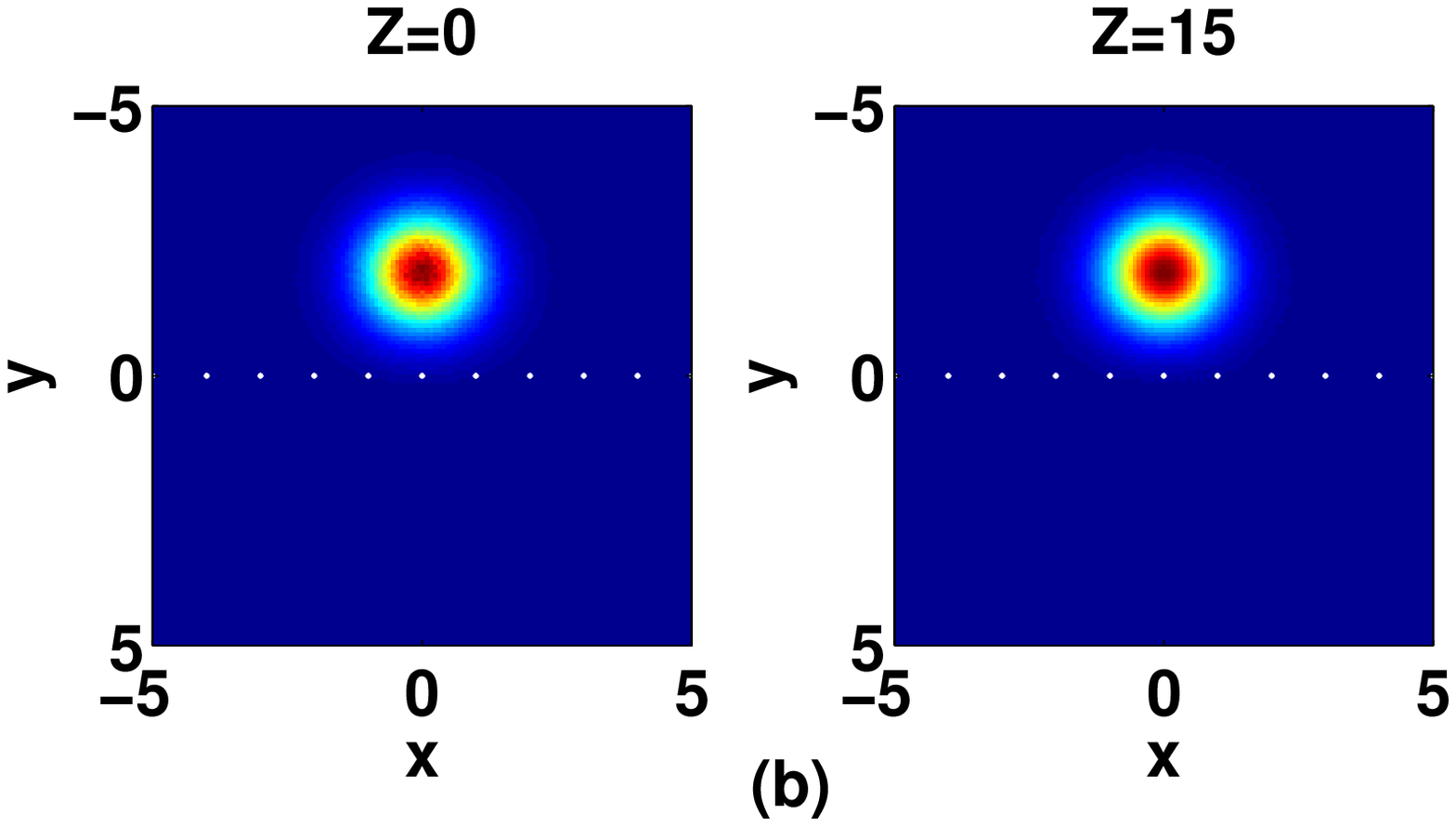}}
{\includegraphics[width=4.5cm,height=4cm]{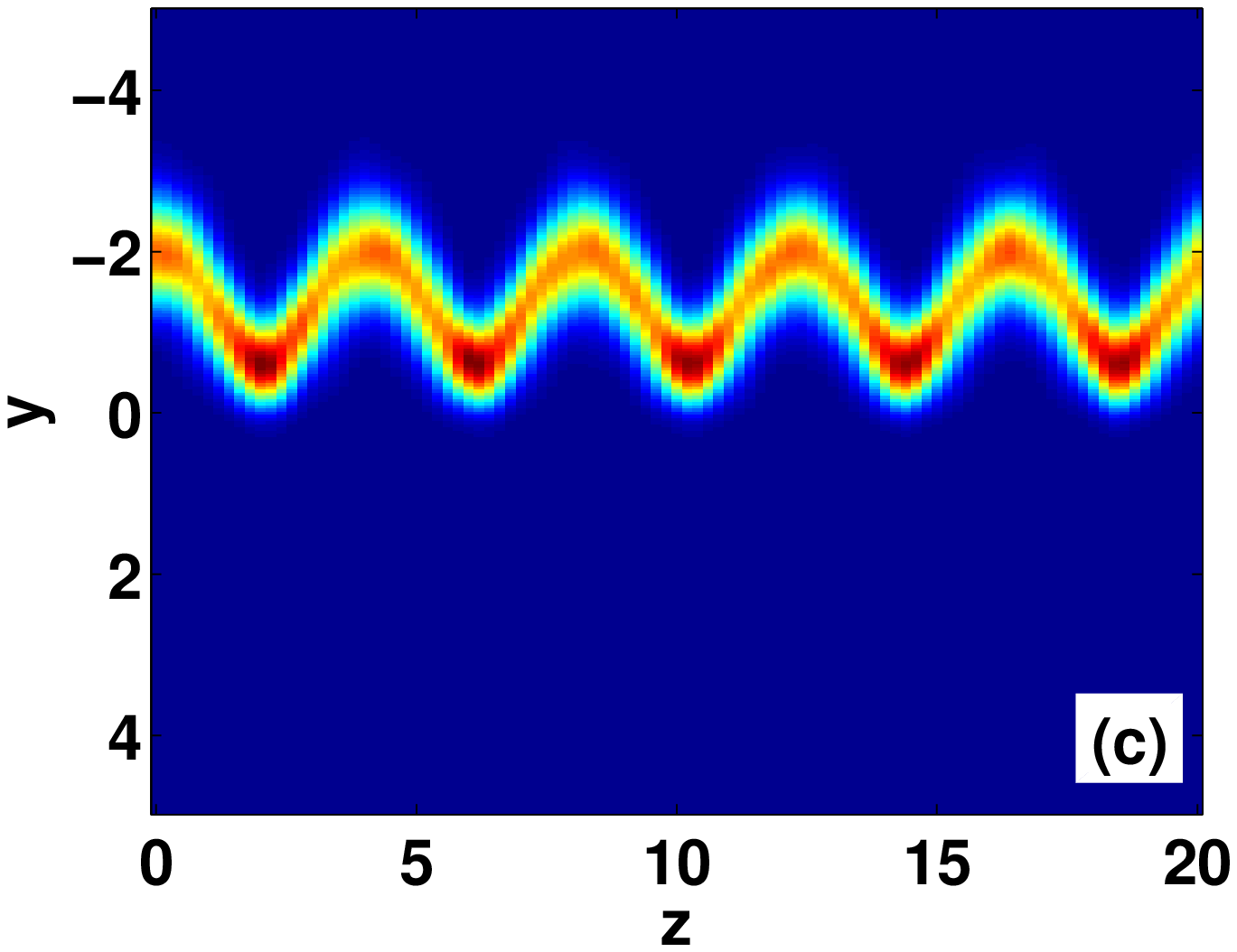}}
\caption{(a) Energy flow $U$ versus the propagation constant $b$ with $d=20$ and $\phi_{d}=7$. (b) Stable propagation of surface
solitons in Fig.1(b)with noise $\sigma^{2}_{noise}=0.05$ for a distance of 15 diffraction lengths. (c) Trajectories of the incident beam with the beam center coordinates $y=-1.99\mu m$. White dashed line indicates interface position. All quantities are plotted in arbitrary dimensionless units.}
\label{fig:two}
\end{figure}

Fig.~\ref{fig:two}(a) shows that the energy flow
$U=\int_{-\infty}^{\infty}\int_{-\infty}^{\infty}|a|^2dxdy$ of the single 2D surface
solitons monotonically increases with $b$ where $dU/db>0$. This shows that the solitons are stable~\cite{ref32,ref33}. Here, to further elucidate the stability of the
surface solitons, we do the direct numerical simulations of
Eqs.~(\ref{eq:three}) and (\ref{eq:four}) with input conditions
$a|_{Z=0}=u(1+\rho)$, where $\rho(x,y)$ is a broadband random
perturbation. The fact which is shown in Fig.~\ref{fig:two}(b) confirms the result of Fig.~\ref{fig:two}(a). Then
we proceed to address the dynamics behavior of the propagation of surface solitons. For convenience, the (1+1)D circumstance is considered.
Fig.~\ref{fig:two}(c) depicts that a narrow beam is launched $y=-1.99\mu m$ away from the interface. The beam maintains a localized shape. However,
it is oscillation in a fully periodic fashion in the virtue of the cooperation of the forces exerted by the boundary and the nonlocal nonlinearity, even if the launch position is far away from the opsition of the surface soliton~\cite{ref33}.

Next, we consider that new soliton solutions at the corner of the two 2D
media. Here, we assume that the normalized
nonlinear contribution to the refractive index at
$x=0$ and $y=0$ have a initial value $\phi=\phi_{d}(\phi_{d}>0)$. The boundary
conditions for the fields at the interface meet the continuity conditions.

\begin{figure}[htbp]
\centering
{\includegraphics[width=4.5cm,height=4cm]{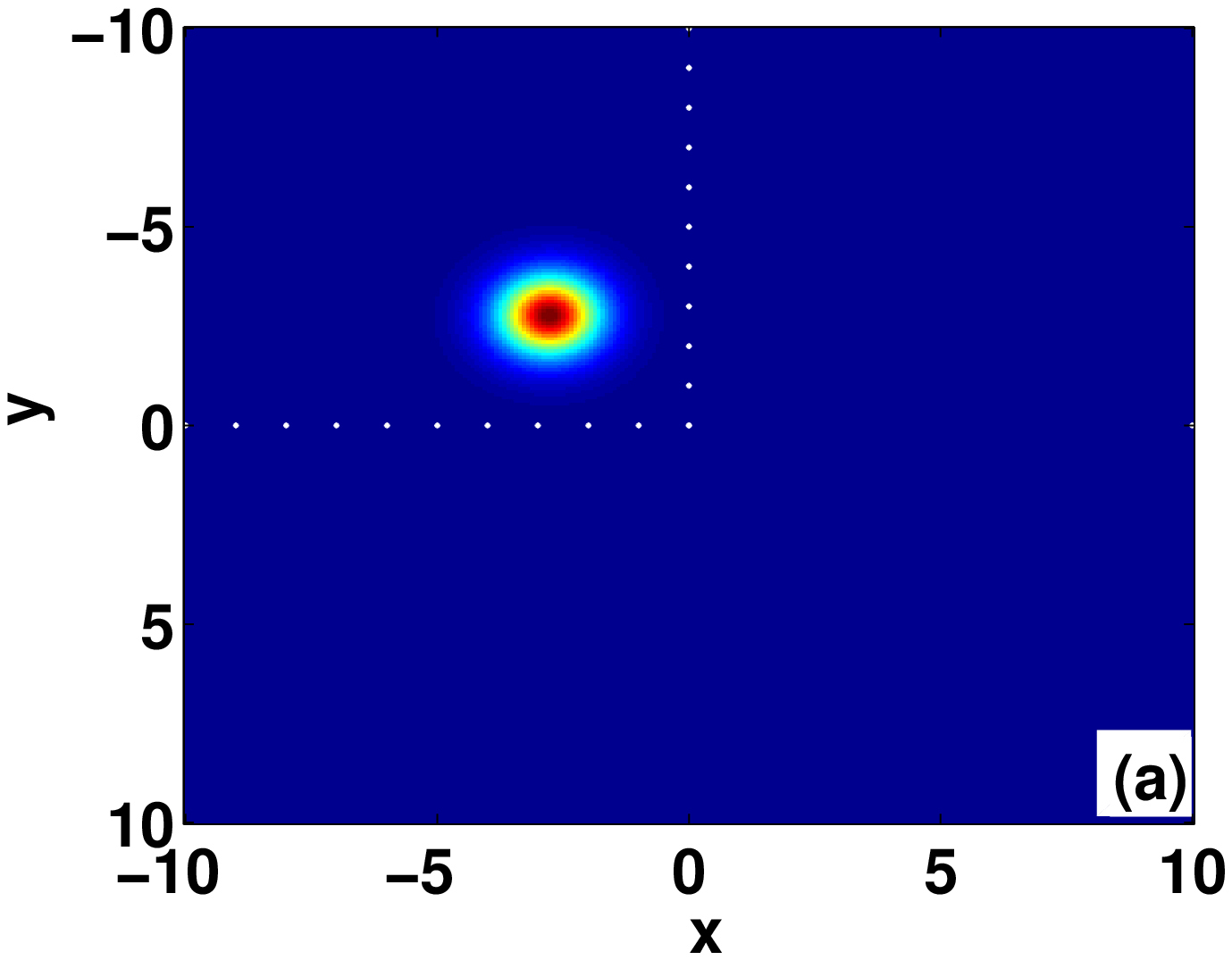}}
{\includegraphics[width=4.5cm,height=4cm]{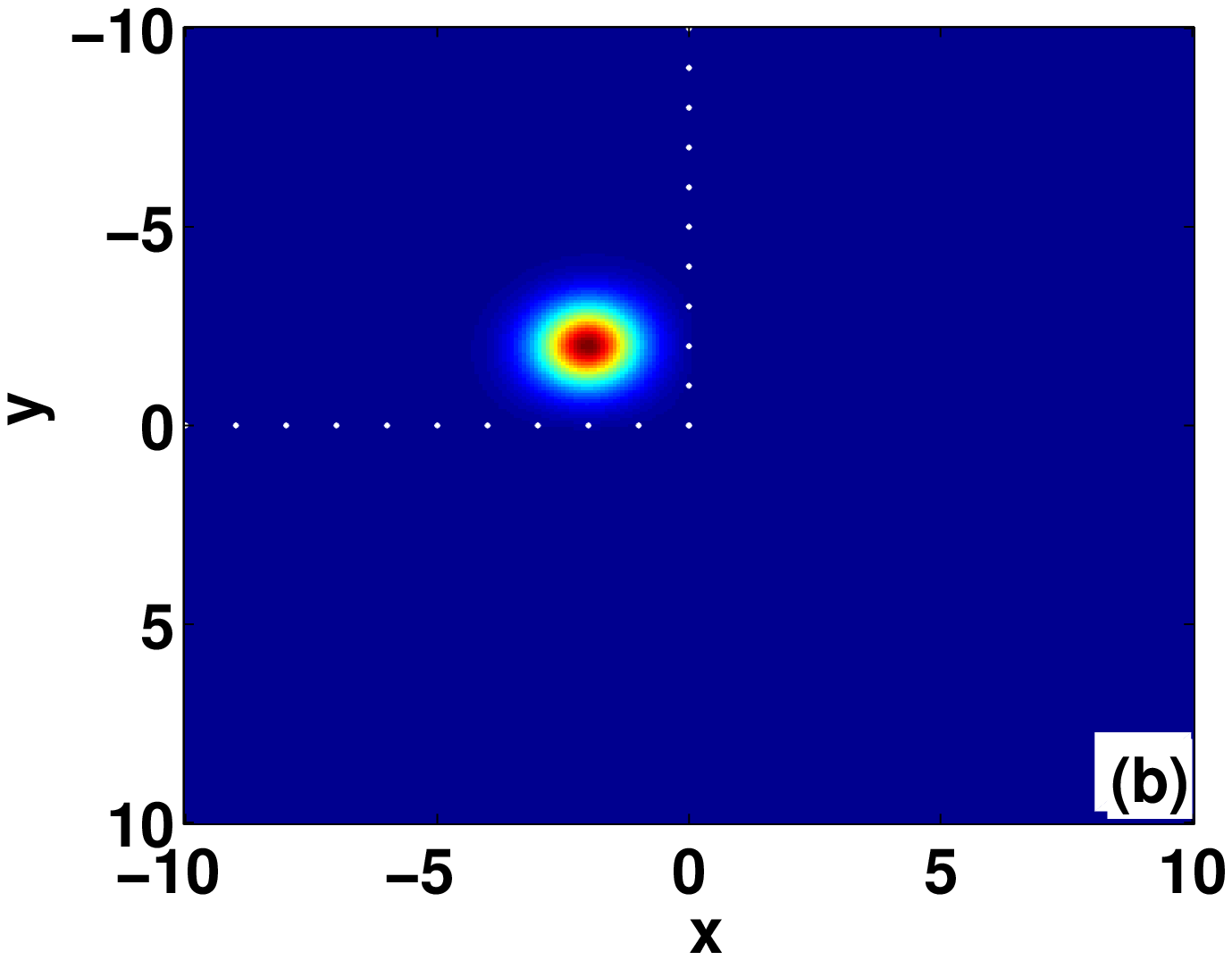}}
{\includegraphics[width=4.5cm,height=4cm]{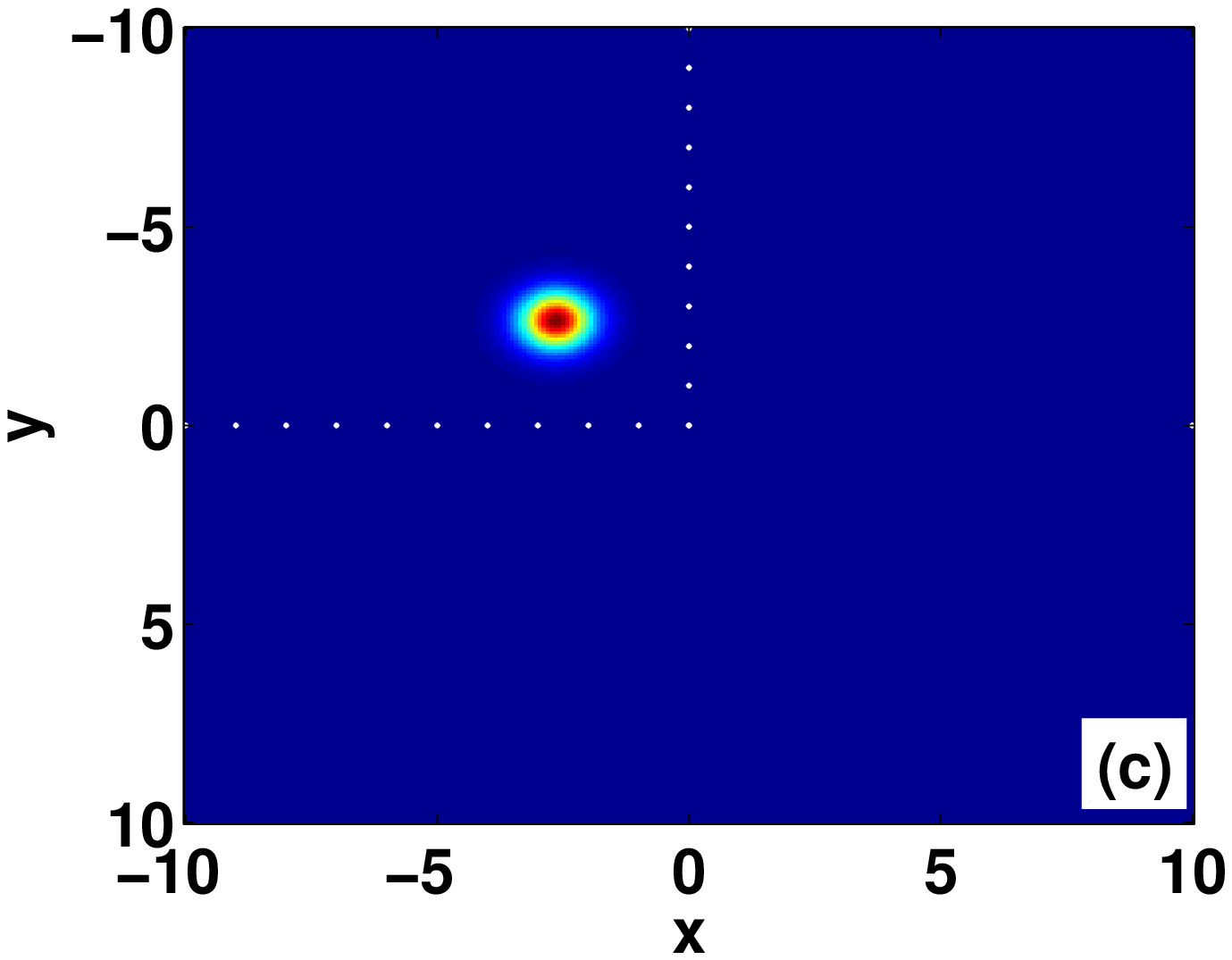}}
{\includegraphics[width=4.5cm,height=4cm]{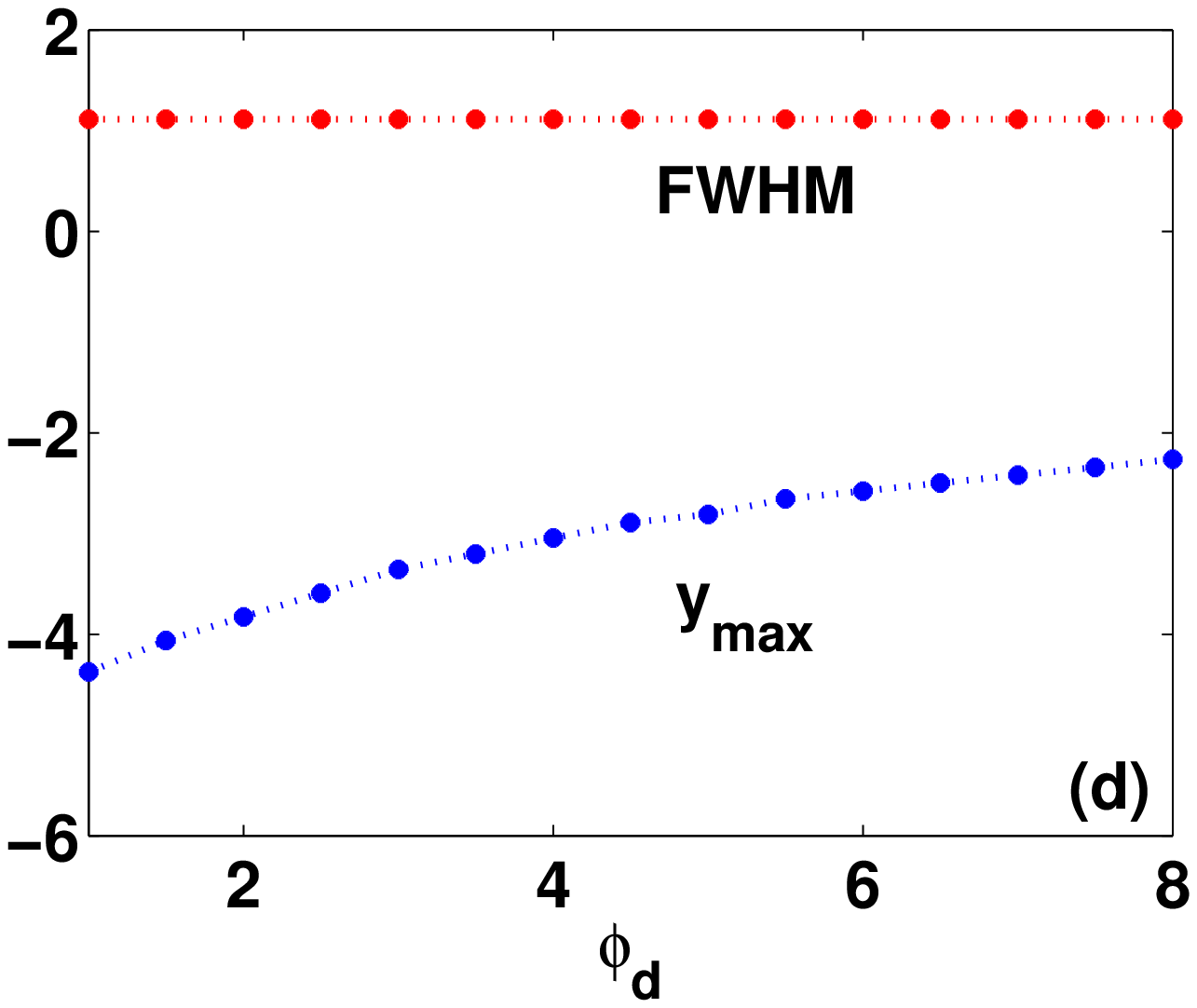}}
{\includegraphics[width=4.5cm,height=4cm]{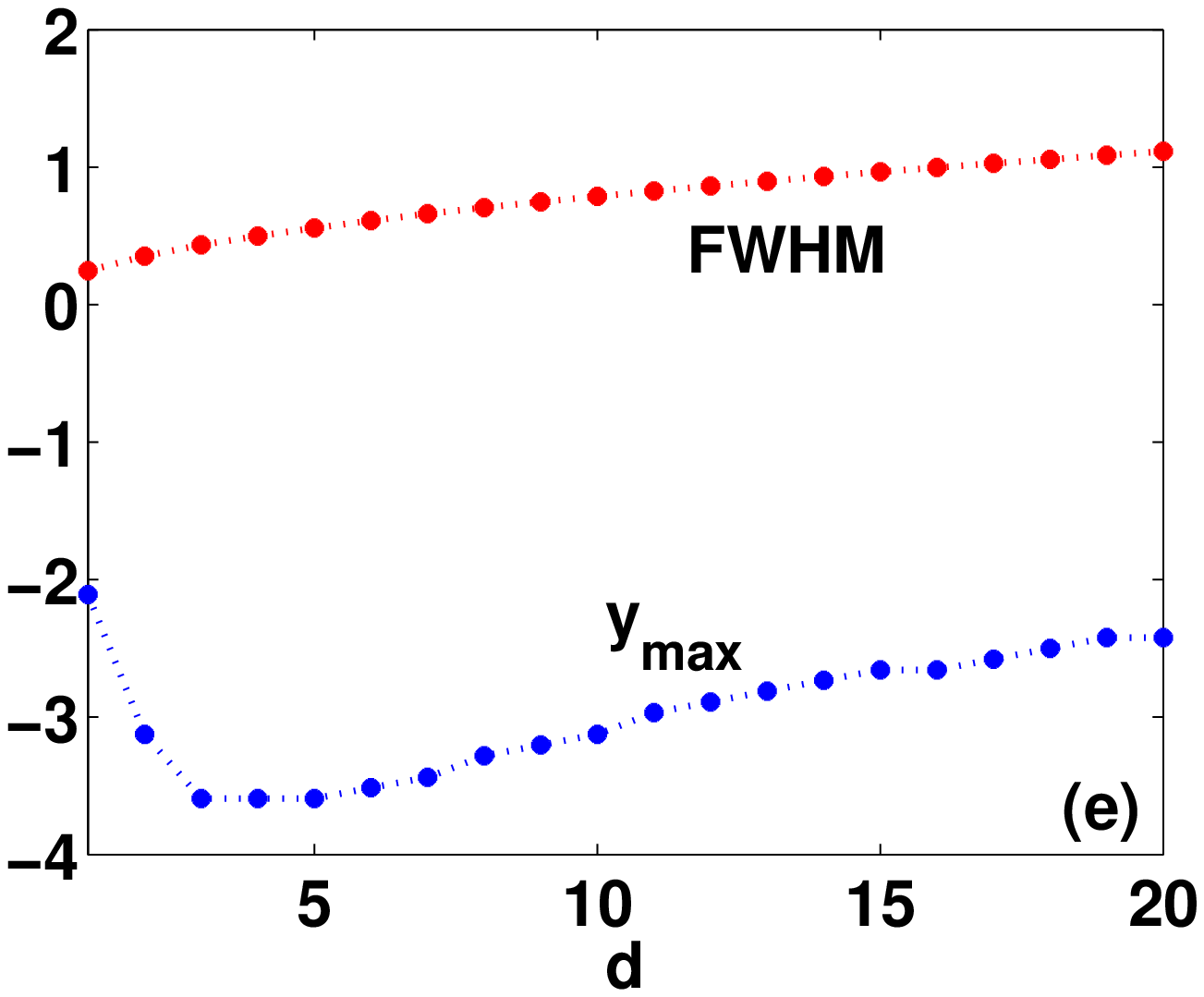}}
\caption{Sketch of 2D single surface solitons at the corner of the interface with (a) $\phi_{d}=5$, $d=20$, (b)
$\phi_{d}=10$, $d=20$ and (c) $\phi_{d}=10$, $d=10$.
The positions of the peak values $y_{max}$ and FWHM versus
(d) the boundary value $\phi_{d}$ and (e) the
nonlocal degree $d$. White dashed line indicates interface position. All quantities are plotted in arbitrary dimensionless units.} \label{fig:three}
\end{figure}

\begin{figure}[htbp]
\centering
{\includegraphics[width=4.5cm,height=4cm]{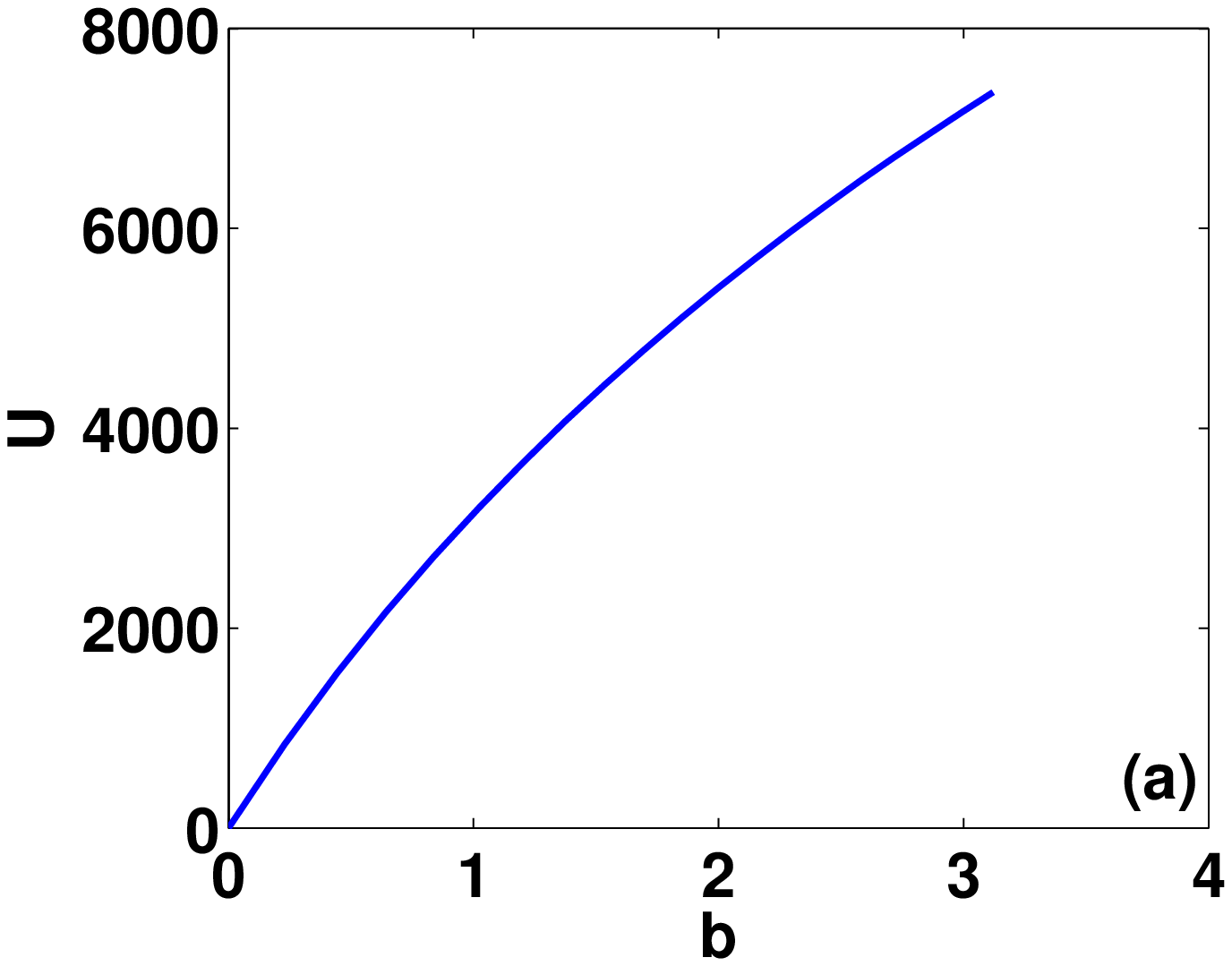}}
{\includegraphics[width=6cm,height=5cm]{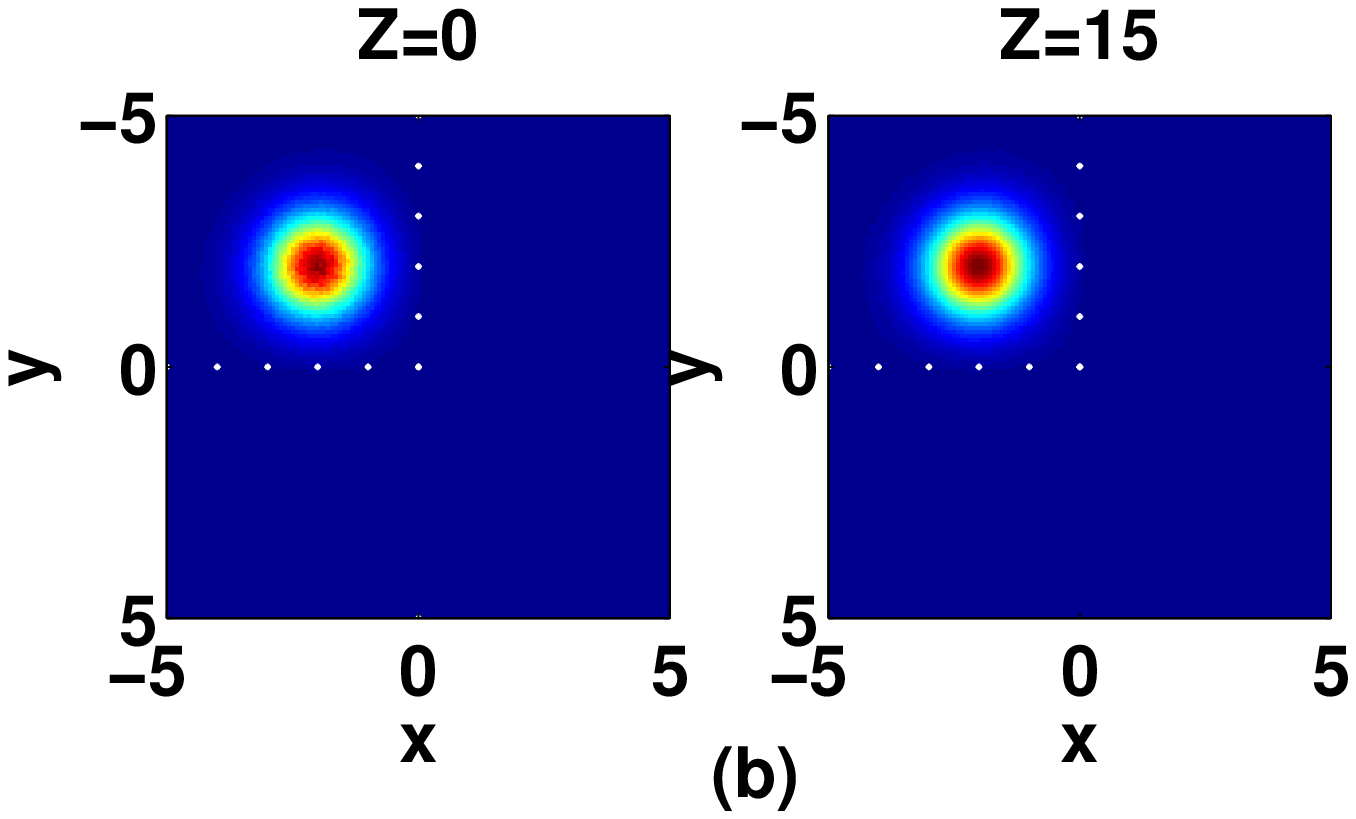}}
\caption{(a) Energy flow $U$ versus the propagation constant $b$ with $d=20$ and $\phi_{d}=10$. (b) Stable propagation of surface
solitons in Fig.3(b) with noise $\sigma^{2}_{noise}=0.05$ for a distance of 15 diffraction lengths. White dashed line indicates interface position. All quantities are plotted in arbitrary dimensionless units.}
\label{fig:four}
\end{figure}

Fig.~\ref{fig:three}(a), (b) and (c) separately show three different solitons. The
soliton is closely attached to the interface with $\phi_{d}=10$ and
$d=20$(see Fig.~\ref{fig:three}(b)), but the solitons are farther
detached from the interface with $\phi_{d}=5$ and $d=20$(Fig.~\ref{fig:three}(a)) or $\phi_{d}=10$ and $d=10$(Fig.~\ref{fig:three}(c)). To further illustrate the influence of $\phi_{d}$ and $d$ on the solitons, we display that $y_{max}$ and FWHM versus $\phi_{d}$ in Fig.~\ref{fig:three}(d) or
$d$ in Fig.~\ref{fig:three}(e). The energy flow
$U$ monotonically increasing with $b$ shown in Fig.~\ref{fig:four}(a) and the direct numerical simulations of
Eqs.~(\ref{eq:three}) and (\ref{eq:four}) with noise $\sigma^{2}_{noise}=0.05$ shown in Fig.~\ref{fig:four}(b) explain that the
solitons are stable. The results can be similarly illustrated as the solitons at the edge of the interface.

\subsection{Two dimensions dipole surface solitons}
In addition to single surface solitons, we also find a 2D stationary dipole surface solitons. The surface solitons are found numerically by a standard relaxation method which converges to a stationary solution after some iterations provided that a suitable guess for initial field distribution. The boundary conditions are the same as the single surface solitons at the edge of the interface.

\begin{figure}[htbp]
\centering
{\includegraphics[width=4.5cm,height=4cm]{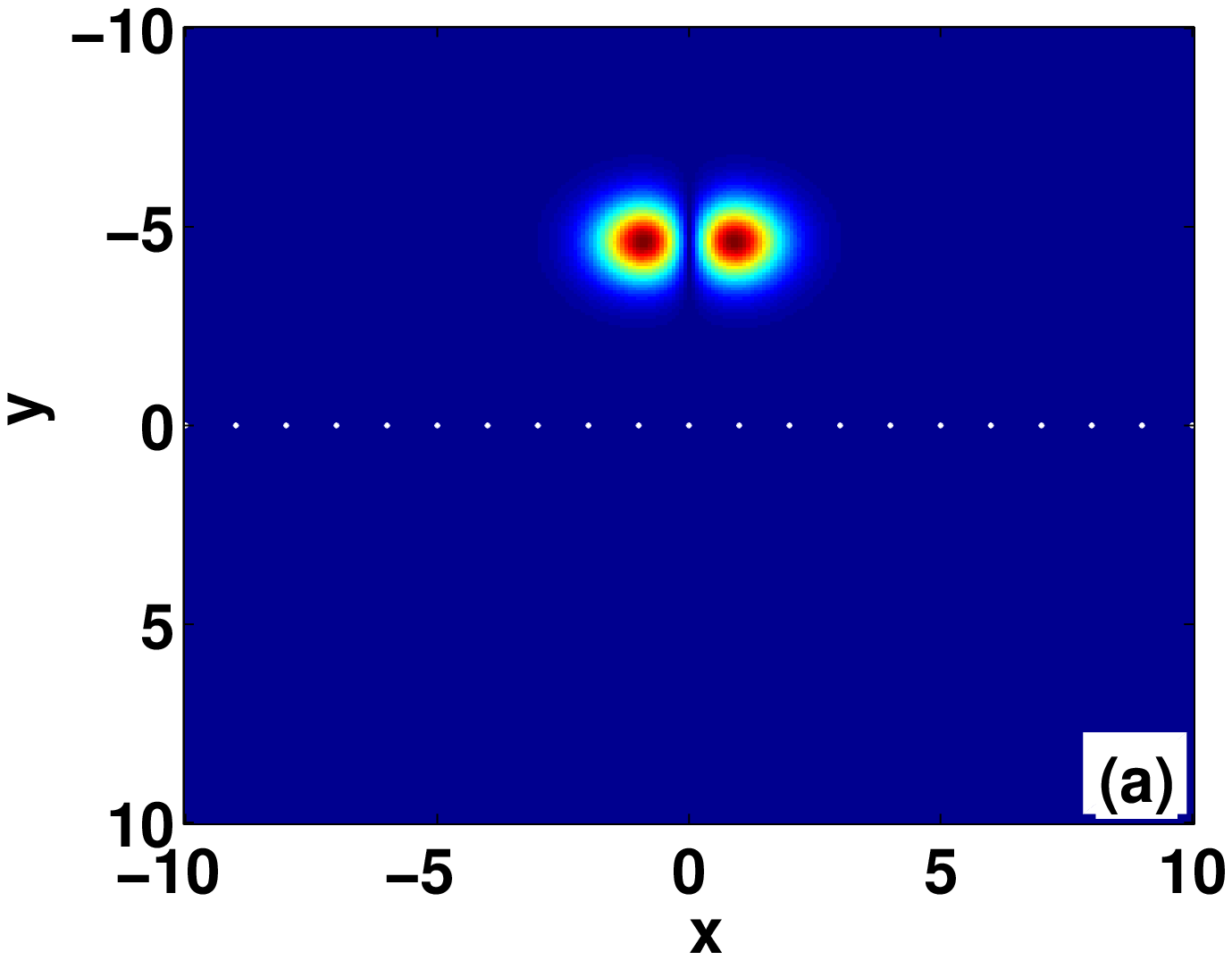}}
{\includegraphics[width=4.5cm,height=4cm]{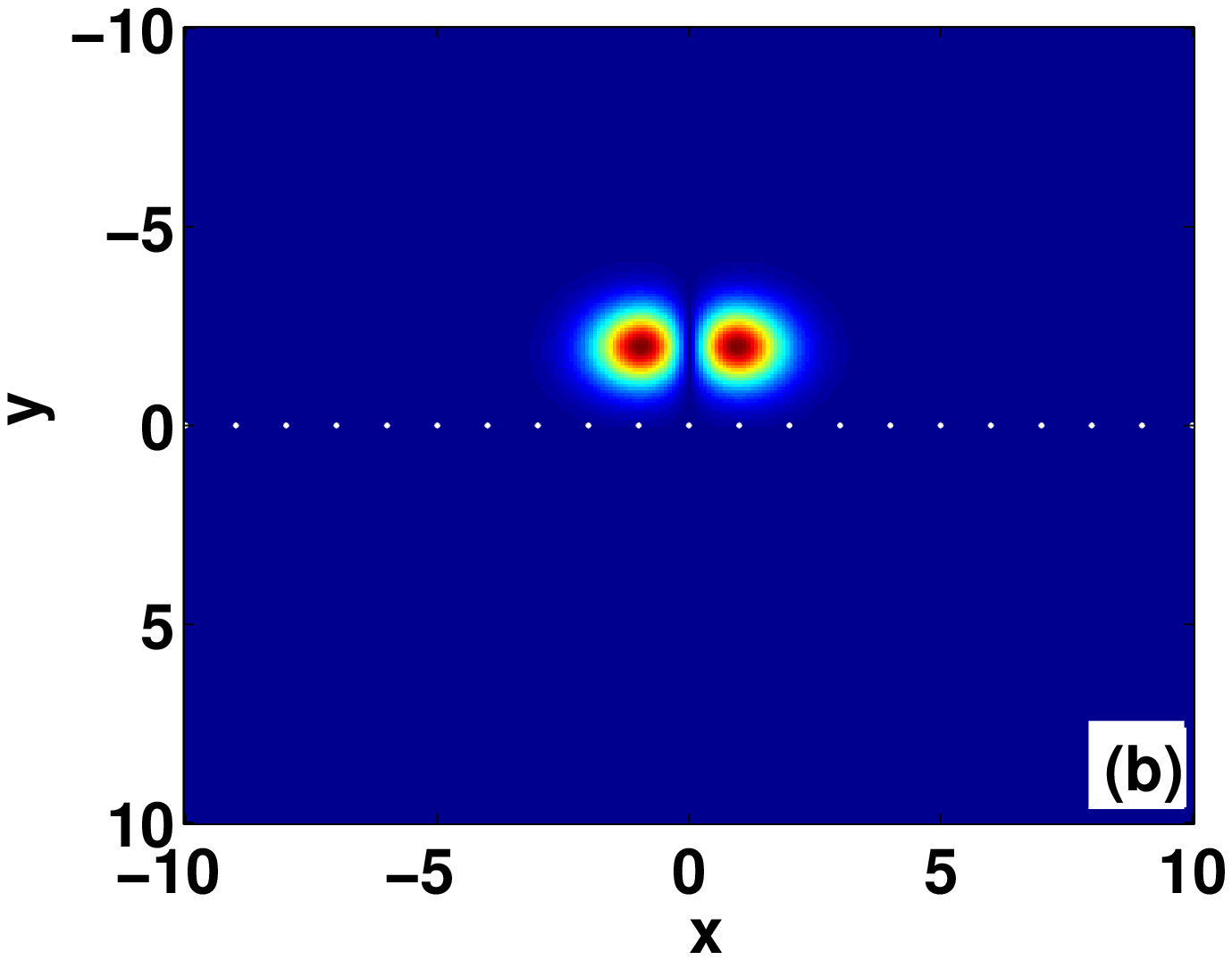}}
{\includegraphics[width=4.5cm,height=4cm]{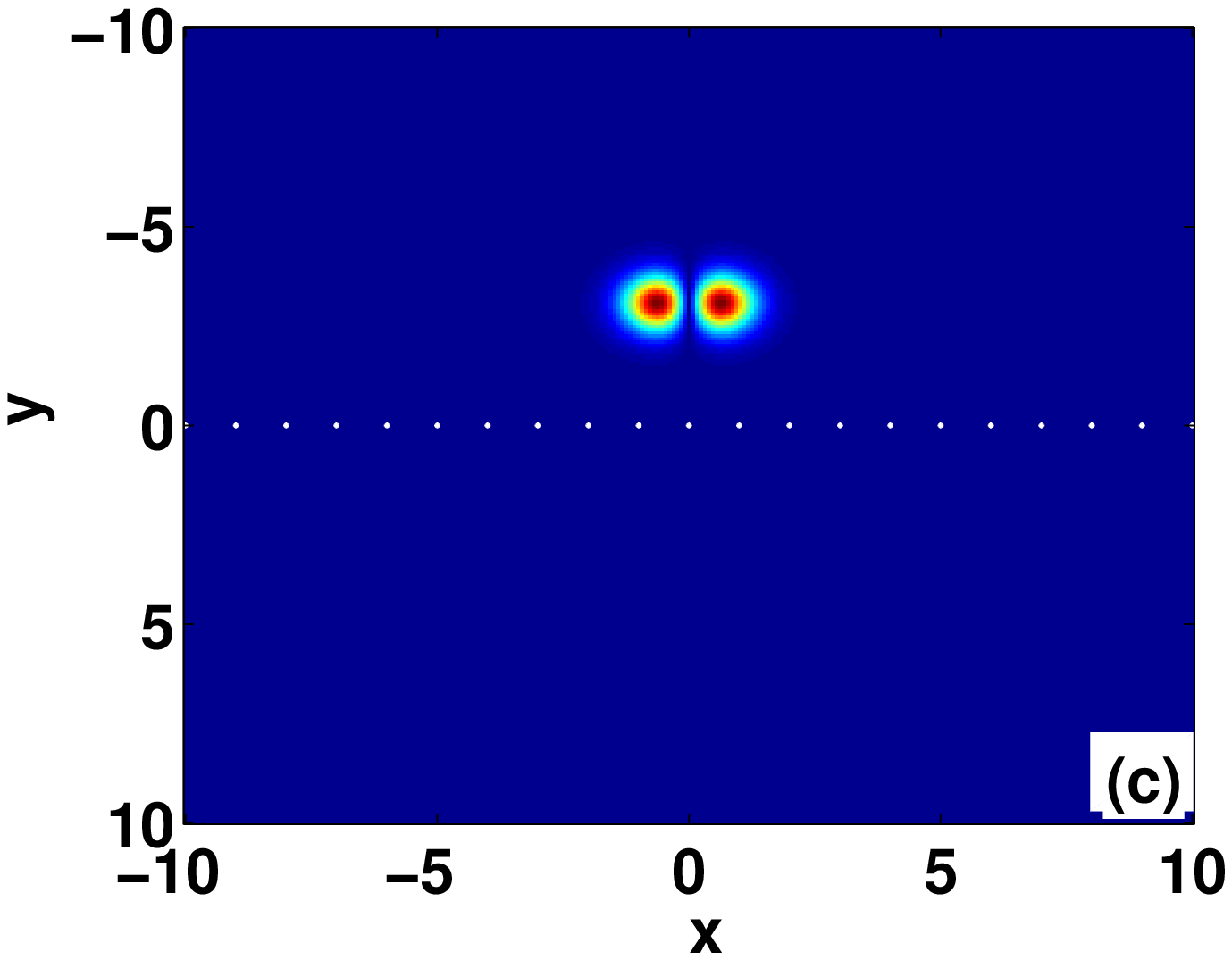}}
{\includegraphics[width=4.5cm,height=4cm]{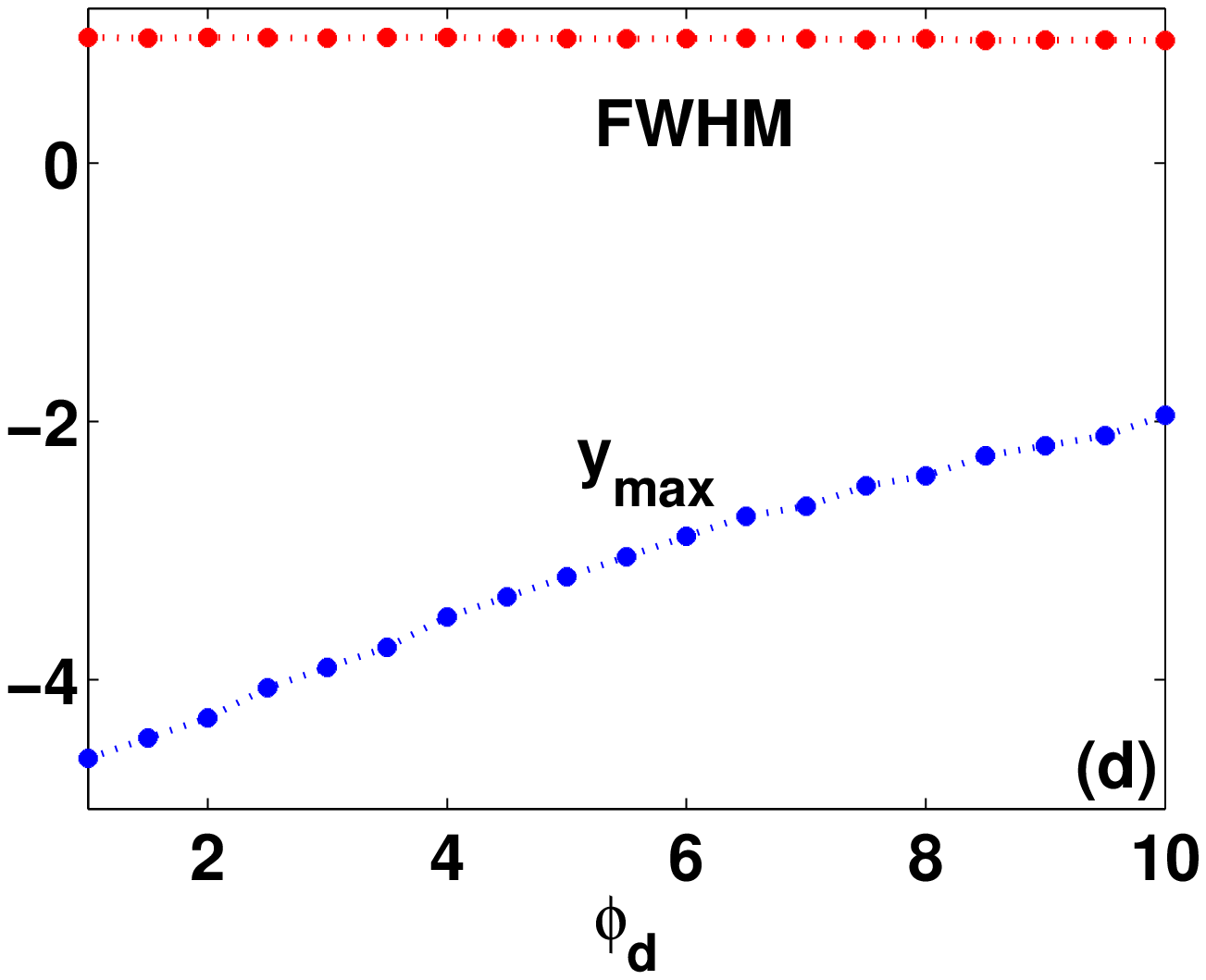}}
{\includegraphics[width=4.5cm,height=4cm]{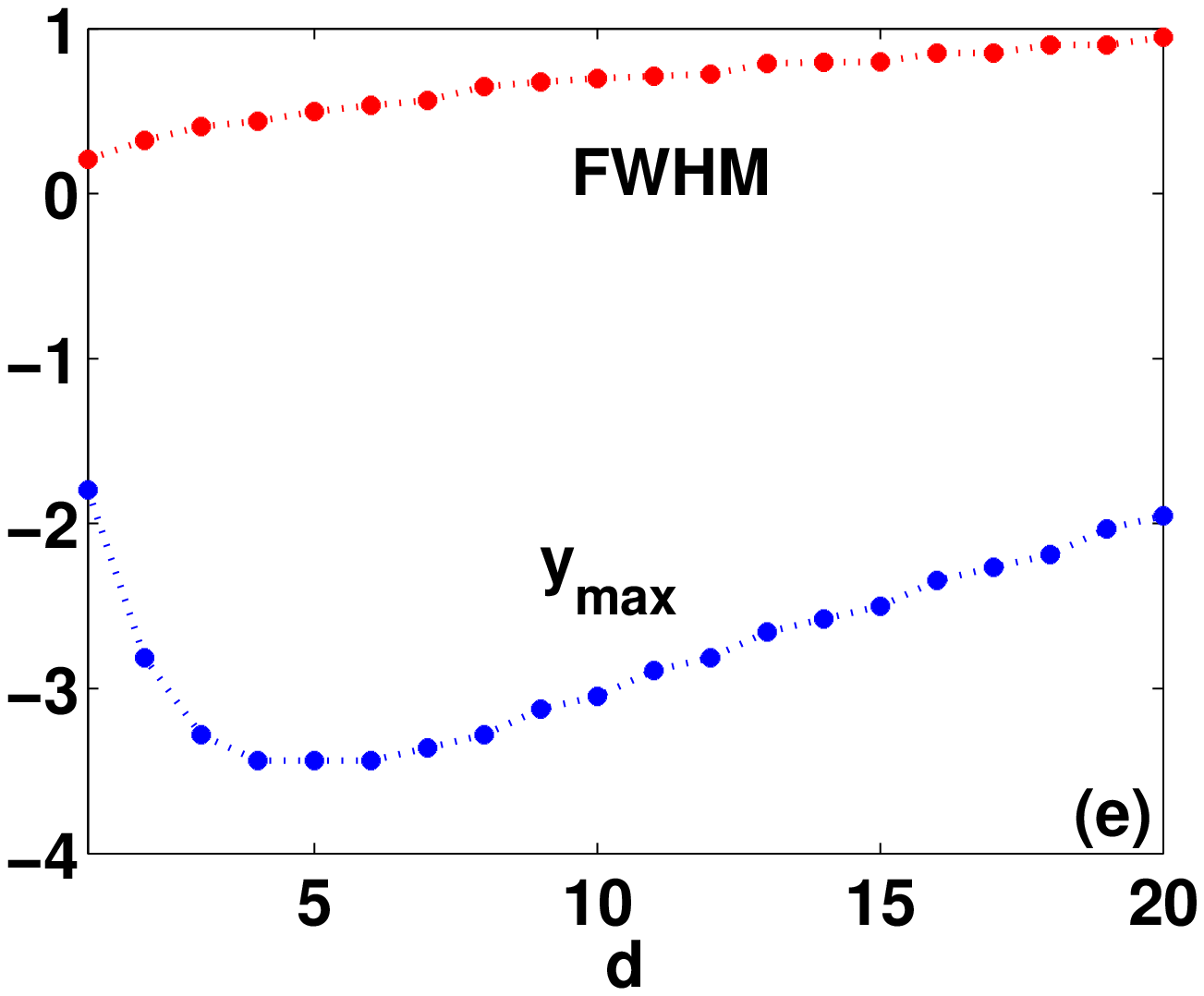}}
\caption{Sketch of 2D dipole surface solitons with (a) $\phi_{d}=1$, $d=20$, (b)
$\phi_{d}=10$, $d=20$ and (c) $\phi_{d}=10$, $d=10$. The positions of the peak values $y_{max}$ and FWHM versus
(d) the boundary value $\phi_{d}$ and (e) the
nonlocal degree $d$. White dashed line indicates interface position. All quantities are plotted in arbitrary dimensionless units.} \label{fig:five}
\end{figure}

Fig.~\ref{fig:five}(b) depicts that the amplitude for a dipole soliton which is closely attached to the interface with $\phi_{d}=10$ and
$d=20$. However, the solitons are farther
detached from the interface with $\phi_{d}=1$ and $d=20$ in
Fig.~\ref{fig:five}(a) or $\phi_{d}=10$ and $d=10$ in
Fig.~\ref{fig:five}(c). Because the poles of solitons are almost symmetric in the $x$ direction, the changing of $y_{max}$ and FWHM of one pole of the solitons can explain the changing of the position and shape of the dipoles. Fig.~\ref{fig:five}(d) and (e) shows $y_{max}$ and FWHM as functions of the boundary value $\phi_{d}$ and the degree
of nonlocality $d$, respectively.

\begin{figure}[htbp]
\centering
{\includegraphics[width=4.5cm,height=4cm]{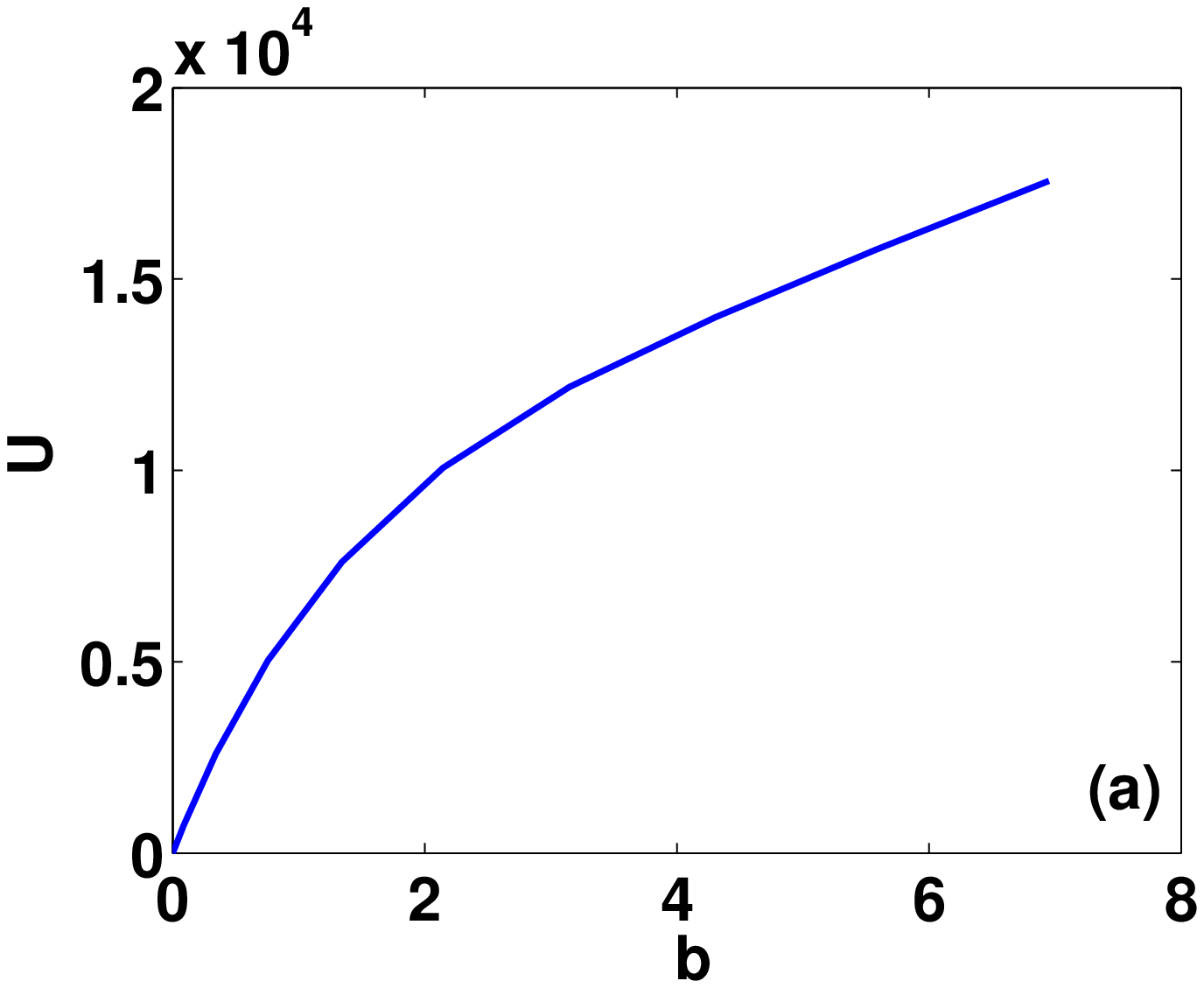}}
{\includegraphics[width=6cm,height=5cm]{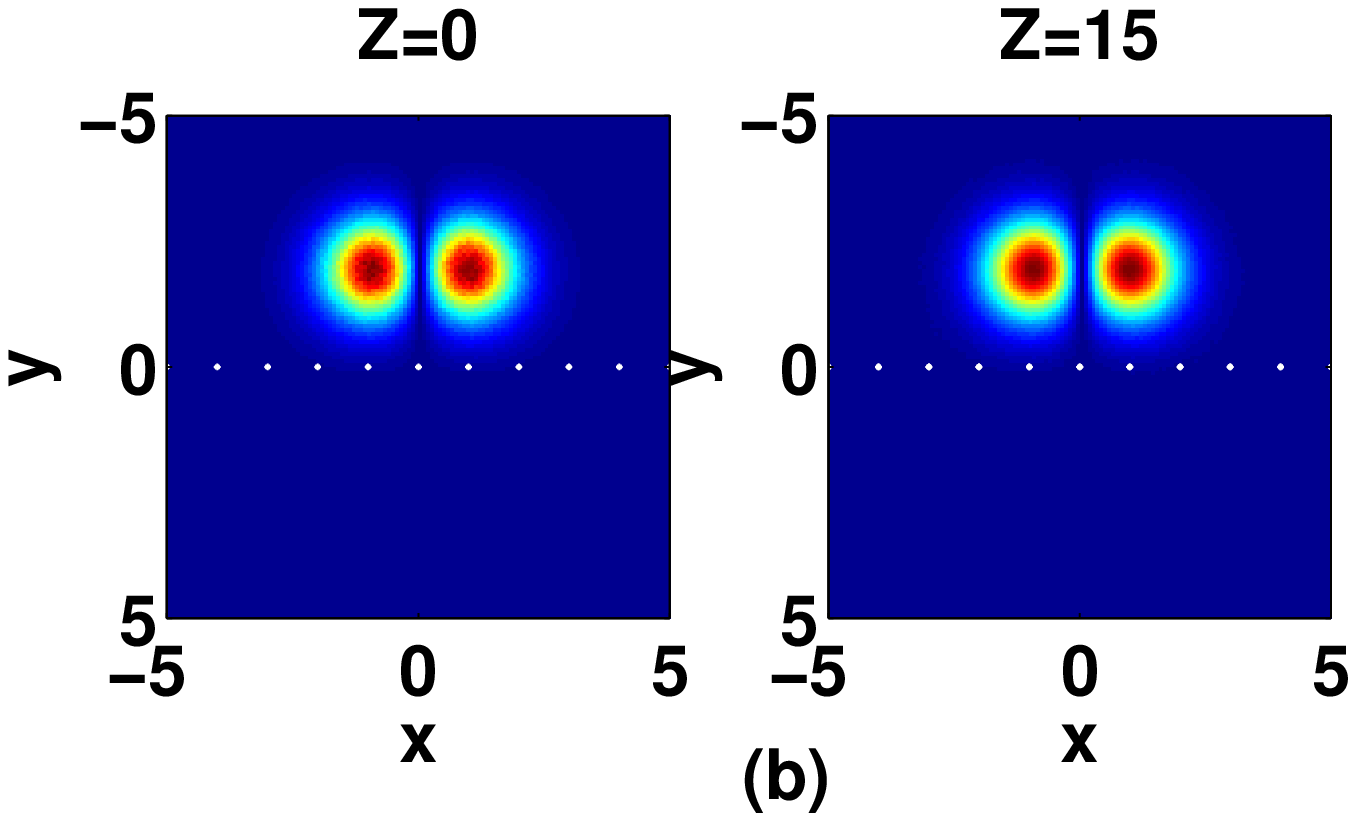}}
{\includegraphics[width=4.5cm,height=4cm]{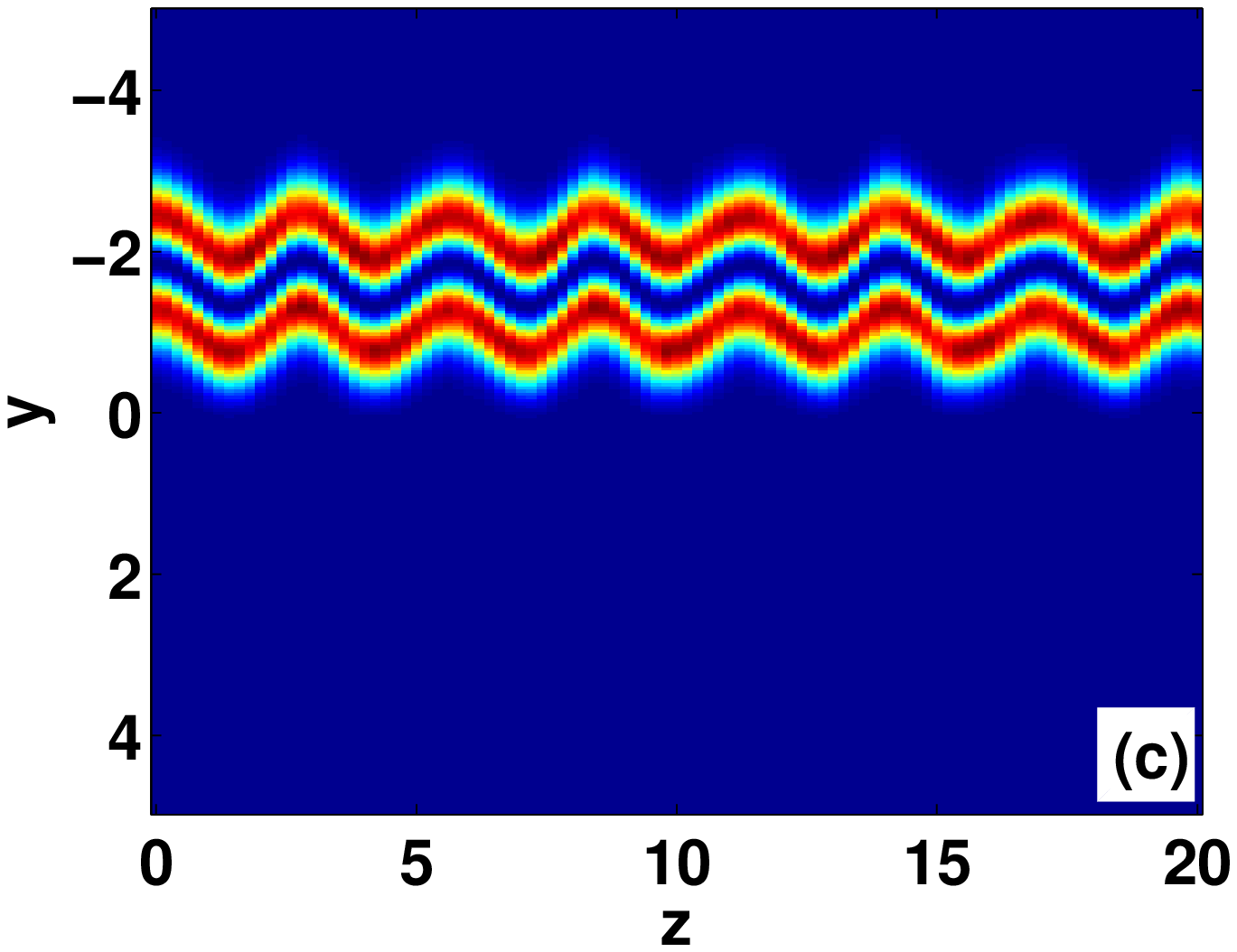}}
\caption{(a) Energy flow $U$ versus the propagation constant $b$ with $d=20$ and $\phi_{d}=10$. (b) Stable propagation of surface
solitons in Fig.5(b)with noise $\sigma^{2}_{noise}=0.05$ for a distance of 15 diffraction lengths. (c) Trajectories of the incident beam with the beam center coordinates $y=-1.91\mu m$. White dashed line indicates interface position. All quantities are plotted in arbitrary dimensionless units.}
\label{fig:six}
\end{figure}

Fig.~\ref{fig:six}(a) shows that the energy flow monotonically increases with the propagation constant for dipole surface solitons. With increasing energy flow, surface dipoles become more localized, i.e., the distance between poles along the $x$ axis and their widths decrease. To further elucidate the stability of the
surface dipoles, we do the direct numerical simulations of
Eqs.~(\ref{eq:three}) and (\ref{eq:four}) with noise $\sigma^{2}_{noise}=0.05$. Particularly, the complex surface solitons are stable in the entire existence domain. Fig.~\ref{fig:six}(b) depicts a typical evolution dynamic. The considerable input perturbations cannot almost cause oscillations of amplitudes of the two poles forming the
dipole, but the dipoles remain their internal structures over
huge distances. Contrarily, in a bulk diffusive medium, it is known that weak input perturbations
can cause slow but progressively increasing oscillations
of the bright spots forming a dipole, resulting specially
in their slow decay into fundamental solitons~\cite{ref34}. So, we
think that the presence of a interface possessing a initial value for $\phi$
leads to stabilization of dipole solitons. This is further illustrated by the results of Fig.~\ref{fig:six}(c). when the beam is launched $y=-1.91\mu m$ away from the interface, the stationary soliton cannot be formed, but it is periodically oscillating. This are very good description of the boundaries of the role of solitons.

\section{Conclusion}
In conclusion, we have addressed (2+1)D surface solitons occurring at the interface between a linear medium and a nonlocal nonlinear medium whose nonlinear contribution to the refractive index has a initial value at the interface.
We find that there exist stable single and dipole surface solitons which do not exhibit a power threshold. The degree of nonlocality have influence on the positions of the peak value and FWHM of the surface solitons, but the initial value can only influence the positions of the peak values of the surface solitons. In addition, when a laser beam is launched away from the interface, the beam will be periodic oscillations, even if the launch position is far away from the interface.

\section*{Acknowledgments}
This research was supported by the National Natural Science
 Foundation of China (Grants No. 11074080 and 10904041), the Specialized Research Fund for the Doctoral Program
 of Higher Education (Grant No. 20094407110008), and the Natural Science Foundation
of Guangdong Province of China (Grant No. 10151063101000017).

\clearpage

\section*{List of Figure Captions}

\noindent Fig. 1. Sketch of 2D single surface solitons at the edge of the interface with (a) $\phi_{d}=1$, $d=20$, (b)
$\phi_{d}=7$, $d=20$ and (c) $\phi_{d}=7$, $d=5$. The positions of the peak values $y_{max}$ and FWHM versus
(d) the boundary value $\phi_{d}$ and (e) the
nonlocal degree $d$. White dashed line indicates interface position. All quantities are plotted in arbitrary dimensionless units.

\noindent Fig. 2. (a) Energy flow $U$ versus the propagation constant $b$ with $d=20$ and $\phi_{d}=7$. (b) Stable propagation of surface
solitons in Fig.1(b)with noise $\sigma^{2}_{noise}=0.05$ for a distance of 15 diffraction lengths. (c) Trajectories of the incident beam with the beam center coordinates $y=-1.99\mu m$. White dashed line indicates interface position. All quantities are plotted in arbitrary dimensionless units.

\noindent Fig. 3. Sketch of 2D single surface solitons at the corner of the interface with (a) $\phi_{d}=5$, $d=20$, (b)
$\phi_{d}=10$, $d=20$ and (c) $\phi_{d}=10$, $d=10$.
The positions of the peak values $y_{max}$ and FWHM versus
(d) the boundary value $\phi_{d}$ and (e) the
nonlocal degree $d$. White dashed line indicates interface position. All quantities are plotted in arbitrary dimensionless units.

\noindent Fig. 4. (a) Energy flow $U$ versus the propagation constant $b$ with $d=20$ and $\phi_{d}=10$. (b) Stable propagation of surface
solitons in Fig.3(b) with noise $\sigma^{2}_{noise}=0.05$ for a distance of 15 diffraction lengths. White dashed line indicates interface position. All quantities are plotted in arbitrary dimensionless units.

\noindent Fig. 5. Sketch of 2D dipole surface solitons with (a) $\phi_{d}=1$, $d=20$, (b)
$\phi_{d}=10$, $d=20$ and (c) $\phi_{d}=10$, $d=10$. The positions of the peak values $y_{max}$ and FWHM versus
(d) the boundary value $\phi_{d}$ and (e) the
nonlocal degree $d$. White dashed line indicates interface position. All quantities are plotted in arbitrary dimensionless units.

\noindent Fig. 6. (a) Energy flow $U$ versus the propagation constant $b$ with $d=20$ and $\phi_{d}=10$. (b) Stable propagation of surface
solitons in Fig.5(b)with noise $\sigma^{2}_{noise}=0.05$ for a distance of 15 diffraction lengths. (c) Trajectories of the incident beam with the beam center coordinates $y=-1.91\mu m$. White dashed line indicates interface position. All quantities are plotted in arbitrary dimensionless units.

%\listoffigures

\clearpage

%% sample sizing command; other sizing commands (and graphics packages) may be used as well

\clearpage

\clearpage

%\clearpage

\end{document}